\documentclass[showpacs,preprintnumbers,amsmath,amssymb,showkeys]{revtex4}
\pdfoutput=1
\usepackage{amssymb}
\usepackage{amsmath}
\usepackage{graphicx}
\usepackage{url,color}
\usepackage[pdftex,colorlinks]{hyperref}
\definecolor{darkblue}{cmyk}{1,0,0,0.8}
\definecolor{darkred}{cmyk}{0,1,0,0.7}
\hypersetup{anchorcolor=black,
  citecolor=darkblue, filecolor=darkblue,
  menucolor=darkblue,pagecolor=darkblue,urlcolor=darkblue,linkcolor=darkblue}
\allowdisplaybreaks

\begin{document}
\author{Matthias Wolfrum}
\affiliation{Weierstrass Institute, Mohrenstrasse 39,
10117 Berlin,
Germany}
\author{Oleh E. Omel'chenko}
\affiliation{Weierstrass Institute, Mohrenstrasse 39,
10117 Berlin,
Germany}
\author{Jan Sieber}
\affiliation{College of Engineering, Mathematics and Physical Sciences,
University of Exeter, North Park Road, Exeter EX4 4QF, United Kingdom}
\title{Regular and irregular patterns of self-localized excitation in arrays of coupled phase oscillators}

\begin{abstract}
  We study a system of phase oscillators with nonlocal coupling in a ring
  that supports self-organized patterns of coherence and incoherence, called chimera states.
  Introducing a global feedback loop, connecting the phase lag to the order parameter,
  we can observe chimera states also for systems with a small number of oscillators.
  Numerical simulations show a huge variety of regular and irregular patterns
  composed of localized phase slipping events of single oscillators.
  Using methods of classical finite dimensional chaos and bifurcation theory,
  we can identify the emergence of chaotic chimera states
  as a result of transitions to chaos via period doubling cascades, torus breakup, and intermittency.
  We can explain the observed phenomena by a mechanism
  of self-modulated excitability in a discrete excitable medium.
\end{abstract}

    % PACS, the Physics and Astronomy Classification Scheme.
                             

\maketitle

\newpage
{\bf Chimera states are self-organized patterns of coherence and incoherence
that appear spontaneously in spatially extended systems of identical oscillators with homogeneous coupling.
After their discovery by Kuramoto and Battogtokh~\cite{kb2002}
they have been investigated mainly in the context of statistical mechanics
using the continuum limit with the number of oscillators tending to infinity,
where they can be described as non-homogeneous equilibrium profiles
of macroscopic (averaged) quantities.
However, as soon as the numbers of oscillators becomes too small
the classical chimera states in the Kuramoto-Sakaguchi oscillators
with nonlocal coupling become unstable and collapse toward a completely coherent state.
This has been explained by characterizing them
as chaotic transients with a lifetime that increases exponentially with the system size~\cite{wo2011}.

As discovered in~\cite{sow2014}, slightly modifying the coupling scheme
with a global feedback on the phase-lag parameter  drastically enhances
the stability of chimera states without otherwise significantly changing them.
For an appropriate choice of the feedback parameters
they appear to be the only attractor in this system
and can be found as stable objects close to the completely coherent state.
Moreover, they can be traced down to very small system size.
This offers the opportunity to study the resulting dynamical regimes
by the methods of classical finite dimensional chaos and bifurcation theory.
Pursuing this approach, we show that chimera states,
which have been described in large systems as a single, statistically stationary regime,
in small systems transform into a huge variety of regular or irregular self-localized patterns.
The variety of different patterns is organized in a complex bifurcation scenario
including transitions  between regular dynamics and chaos
by period-doubling cascades, torus breakup, and intermittency.
}

\section{Introduction}
The self-organized formation of patterns in homogeneous media is a fundamental paradigm in nonlinear science.
Recently, a lot of interest has been attracted by self-organized coherence-incoherence patterns
in spatially extended oscillator systems, called {\em chimera states}~\cite{as2004,pa2015}.
After their first description by Kuramoto and Battogtokh in a system of coupled phase oscillators,
similar dynamical regimes have been reported
for a variety of theoretical and experimental settings, including e.g. chemical, electronic
and mechanical oscillators~\cite{tns2012,wk2013,wk2014,sskg-m2014,hmrhos2012,rrhsg2014,bzsl2015,mtfh2013}
as well as models of neuronal systems~\cite{lc2001,oohs2013,pm2014,l2014}.
In all of these studies, a basic requirement for the observation of chimera states
was a sufficiently large number of oscillators of typically at least $N\approx 40$.
This imposes a substantial restriction on experimental realizations, see~\cite{tns2012},
and leads to the fact that up to now, with a few exceptions~\cite{gbcfmf2014,ab2015,bzsl2015},
the main tool for theoretical investigations of chimera states
has been the continuum limit with the number of oscillators tending to infinity.
In this paper, we will study a slightly modified system,
which has been recently presented in~\cite{sow2014},
where an additional global feedback stabilizes the chimera states
in a way such that they can be observed for a smaller number of oscillators.

As in the classical chimera system, we start with~$N$ identical Kuramoto-Sakaguchi phase oscillators of the form
\begin{equation}\label{eq:1}
\frac{d \theta_k}{d t} = \omega - \frac{2\pi}{N} \sum\limits_{j=1}^N G_{kj} \sin( \theta_k - \theta_j + \alpha)
\mbox{,}\quad k=1,\ldots,N,
\end{equation}
with phases $\theta_k\in[0,2\pi)$ and a coupling matrix~$G\in \mathbb{R}^{N\times N}$.
The natural frequencies of the oscillators are all chosen identical,
such that without loss of generality we can assume $\omega=0$.
We consider a one-dimensional array of oscillators, where each oscillator is located
at the position $x_k=2k\pi/N-\pi$ in the interval $[-\pi,\pi]$
and which is closed by periodic boundary conditions.
Then, the coupling constants~$G_{kj}$ can be expressed by a coupling function~$G(x)$,
depending on the distances $x=x_k - x_j$ such that the coupling is smaller between more distant oscillators.
In this paper we choose a sinusoidal function,
\begin{equation}\label{eq:2}
G_{kj}=G(x_k - x_j) = \frac{1}{2\pi}[1+A\cos(x_k - x_j)]\mbox{,}
\end{equation}
as it has been suggested in~\cite{as2004}. In order to find a classical chimera state in this system,
the phase-lag parameter~$\alpha$ in the phase response function,
governing the attraction and repulsion between the oscillators,
has to be well tuned to values slightly below~$\pi/2$.
However, following~\cite{sow2014}, we choose the phase lag~$\alpha$ not as a constant parameter.
Instead, we introduce a global feedback loop
\begin{equation}\label{eq:3}
 \alpha(t)=\frac{\pi}{2} - K(1-r(t))\mbox{,}
\end{equation}
between phase lag~$\alpha$ and the  global order parameter
\begin{equation}\label{eq:4}
r(t)= \left| \frac{1}{N} \sum\limits_{j=1}^N e^{i \theta_{j}(t)} \right|.
\end{equation}
In~\cite{sow2014} it has been described in detail how this type of global feedback
can be interpreted as a proportional control that stabilizes the chimera states.
For systems with a large number of oscillators it is non-invasive on average,
i.e. the branch of equilibrium solutions of the continuum limit remains unchanged,
while only its stability properties change.
In particular, it has been shown that for suitably chosen control parameters
the completely coherent state loses its stability
and for large~$N$ the chimera state is the only attractor of the system.
The loss of stability of the completely coherent state rules out
the main instability mechanism for chimera states with small~$N$.
Indeed, as shown in~\cite{wo2011} chimera states in the classical Kuramoto-Battogtokh system
with nonlocal coupling show sudden collapses to the completely coherent state
and can be characterized as chaotic transients
with a lifetime that increases exponentially with the system size.
Due to this effect, it is impossible to observe stable chimera states for small~$N$
in the Kuramoto-Battogtokh system without the feedback term.
The main purpose of the present paper is to give a detailed description
of the dynamical phenomena in the feedback system with a small number of oscillators,
using methods of classical finite dimensional chaos and bifurcation theory.
We first present a detailed numerical study of the dynamics
displaying a huge variety of regular and irregular stationary and propagating patterns.
Then, we give an explanation of the observed phenomena in this system,
describing it as a discrete medium at the threshold between stationary excitable and oscillatory behavior,
where the pattern formation is caused by an interplay of activation and inhibition
due to the nonlocal coupling and the global feedback,
which leads to a  self-organized spatial modulation of the excitability threshold.
\begin{figure}[h]
\begin{center}
\includegraphics[width=0.5\textwidth]{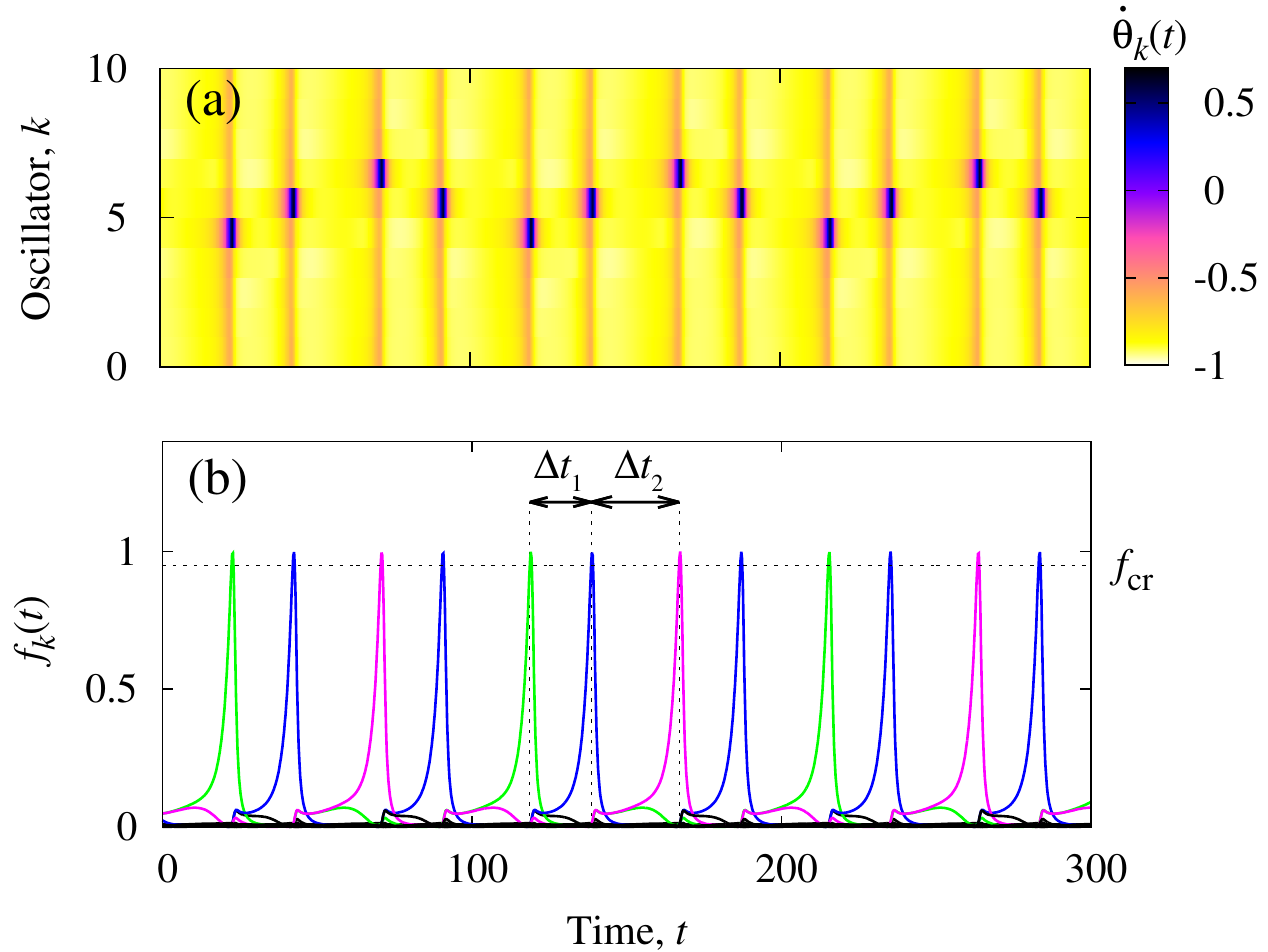}
\end{center}
\caption{(a): Space-time profile of the phase velocities~$\dot\theta_k(t)$
  for $N=10$ oscillators with $A=0.9$ and $K=2.8$
  (dark spots are phase slips, bright regions have coherent motion).
  (b): The peaks of~$f_k(t)$ (which is defined in~(\ref{eq:f_k})) indicate the times
  at which the $k$-th oscillator is in anti-phase to the respective local mean field
  ($k=5$ is green, $k=6$ blue, and $k=7$ magenta). For all other values of $k$,
  $f_k(t)$ remains close to zero (black curve).
  Vertical lines indicate times at which events~$t_s$ are recorded.
}
\label{Fig:tprof}
\end{figure}

\section{Symmetries of the system}
System~(\ref{eq:1})--(\ref{eq:4}) possesses several symmetries.
It is equivariant with respect to phase shifts,
i.e. adding a constant to a given solution $\theta_k(t),\,k=1\ldots N$ provides again a solution of the system.
In particular, this leads to the possibility of uniformly rotating periodic solutions of the form
$$
\theta_k(t) = \Omega t + \hat\theta_k\:\:\: \mathrm{mod}\:\:\: 2\pi,
$$
so called relative equilibria that turn into  equilibria in a corotating coordinate frame.
In addition there can appear relative periodic solutions,
i.e. quasiperiodic solutions that turn into  periodic solutions
in a corotating coordinate frame with a suitably chosen frequency~$\Omega$.
Moreover, the dihedral group~$D_N$ acting on the indices $k=1\ldots N$ obviously maps solution to solutions.
As we will see below, this opens the possibility for relative periodic solutions
with different spatio-temporal symmetries,
i.e. solutions returning after a period $P>0$ not to the initial state,
but to a state related to it by symmetry, i.e.
\begin{equation}\label{eq:RPO}
\theta_k(t+P) = \theta_{\sigma(k)}(t) + \psi\:\:\: \mathrm{mod}\:\:\: 2\pi
\end{equation}
for some group elements $\psi\in S^1,\,\sigma\in D_N$.

\section{Numerically calculated patterns and bifurcation diagrams}

Numerical simulations of system (\ref{eq:1})--(\ref{eq:4}) show a huge variety of periodic and chaotic solutions,
displaying different types of  stationary and traveling patterns.
Fig.~\ref{Fig:tprof}(a) shows an example of a numerically obtained time profile
for~$\dot\theta_k(t)$ for $N=10$ oscillators.
Note that this solution displays clearly separated phase slipping events,
which appear in the phase velocity plots as sharp peaks
separated by long intervals where the velocity is nearly constant.
At the velocity peaks, the phase~$ \theta_k$ of the corresponding oscillator
is in anti-phase to its local mean field
\begin{equation}\label{eq:mf}
W_k := R_k e^{i \Theta_k} = \frac{2\pi}{N} \sum\limits_{j=1}^N G_{kj} e^{i \theta_j},
\end{equation}
while in the intervals in between all oscillators are nearly in phase with their local mean field.
This leads to the fact that the function
\begin{equation}\label{eq:f_k}
f_k(t) = \frac{1}{2} \left( 1 - \cos(\theta_k(t) - \Theta_k(t)) \right),
\end{equation}
which measures the distance between~$\theta_k$ and the local mean field phase~$\Theta_k$,
has a sharp peak as well, see Fig.~\ref{Fig:tprof}(b).
We use this observation for the construction of a Poincar\'e section in the following way:
We record the time moment~$t_s$ of a velocity peak,
using the condition  that for some oscillator~$k_s$ the function
$
f_{k_s}(t)\arrowvert_{t=t_s}
$
enters into the region above the chosen level $f_{\mathrm{cr}}=0.95$.

\begin{figure}[ht]
\begin{center}
\includegraphics[width=0.98\textwidth]{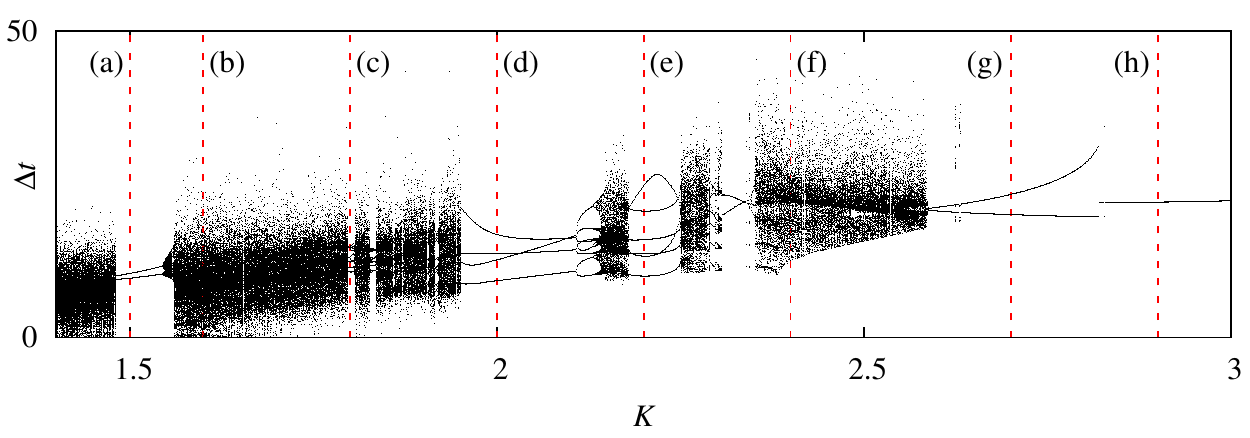}\\[2mm]
\includegraphics[width=0.98\textwidth]{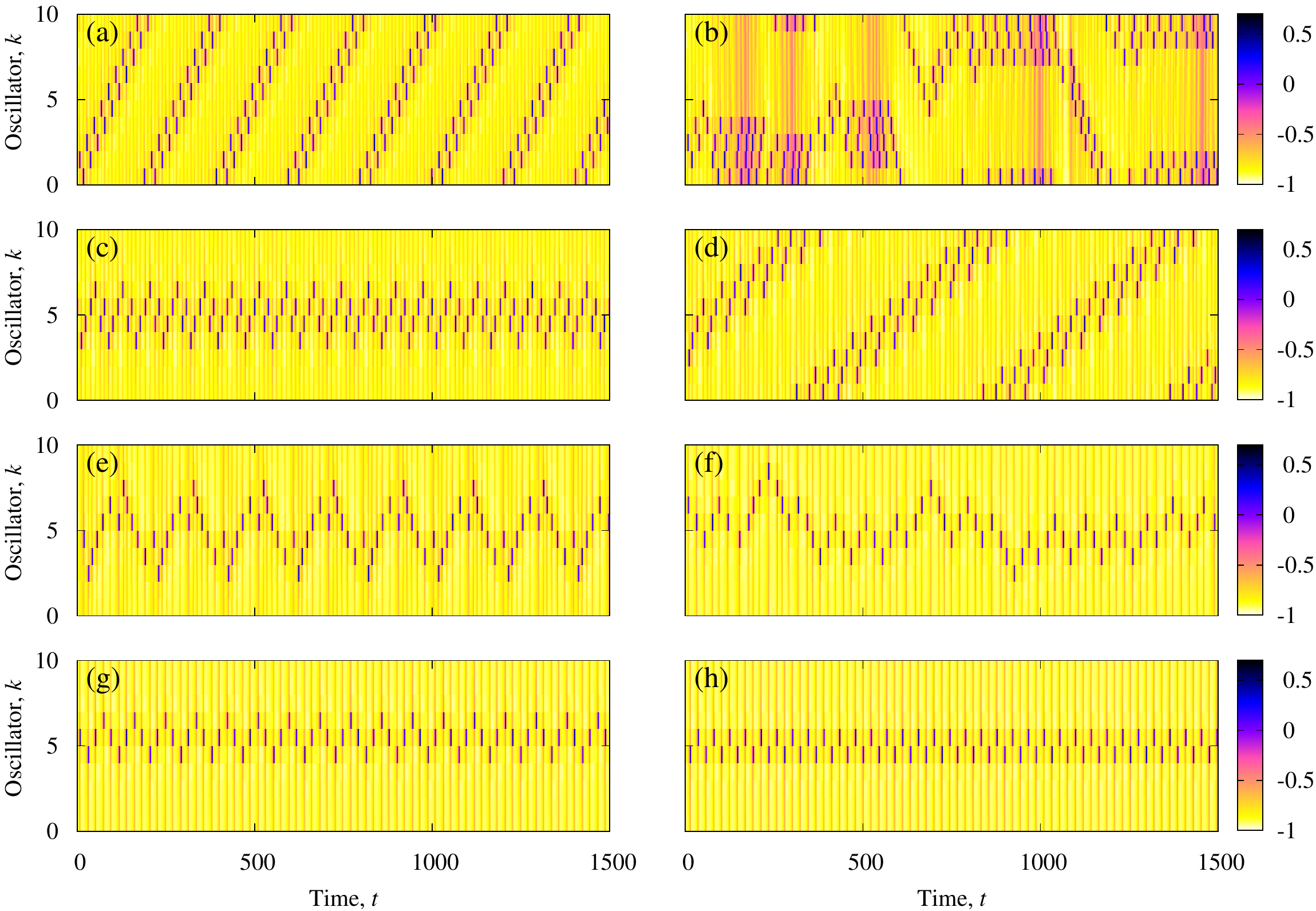}
\end{center}
\caption{Bifurcation diagram: sampled time intervals between velocity peaks for varying~$K$.
Panels~(a)--(h): space time plots of the phase velocities for selected values of~$K$
(dashed vertical lines in the bifurcation diagram).
Parameters: $N = 10$, $A = 0.9$.}
\label{Fig:1}
\end{figure}

In our simulations we have chosen $A=0.9$ and treated~$K$ as a bifurcation parameter.
Note that the feedback term~(\ref{eq:3}) provides an attractive coupling only for $K>0$.
In Fig.~\ref{Fig:1}, we present a bifurcation diagram, taking $K\in [1.4,3]$
where we found the bifurcations and transitions from regular to chaotic dynamics.
We sampled the time intervals $\Delta t_s=t_s-t_{s-1}$ between two consecutive velocity peaks,
starting for each parameter value with a random initial condition
and discarding an interval of $20\,000$ time units for the transients.
The diagram shows several regions with different types of chaotic and regular solutions.
The panels (a)--(h) in Fig.~\ref{Fig:1} show the time traces of the solutions
for selected values of the parameter~$K$, indicated by dashed vertical lines in the bifurcation diagram above.

Note that both the regular and the chaotic solutions are reminiscent
of the chimera states that can be observed in such systems for large values of~$N$.
One can clearly distinguish self-organized regions of
coherent oscillation from those regions where velocity peaks are
localized.  For the regular solutions, these regions are either
stationary, as in (c), (e), (g), (h), or propagating at a constant
speed, as in (a), (d).  For the chaotic solutions (b), (f), one can
also observe regions where velocity peaks are localized, however, they
show an irregular motion of their position in space.
As it has been shown in~\cite{owm2010}, there is a similar irregular motion
also for classical chimera states with larger~$N$.
In the case of~\cite{owm2010}, the motion can be described
as a Brownian process with a diffusivity proportional to~$N^{-2}$.
We study now in more detail the different types of transitions to chaos
that can be observed in this system.
%%%%%%%%%%%%%%%%%%%%%%%%%%%%%%%%%%%%%%
%%%%%%%%%%%%%%%%%%%%%%%%%%%%%%%%%%%%%%%

\subsection{Period doubling cascade}
%%%%%%%%%%%%%%%%%%%%%%%%%%%%%%%%%%%%%%%
\begin{figure}[ht]
\begin{center}
\includegraphics[width=8.5cm]{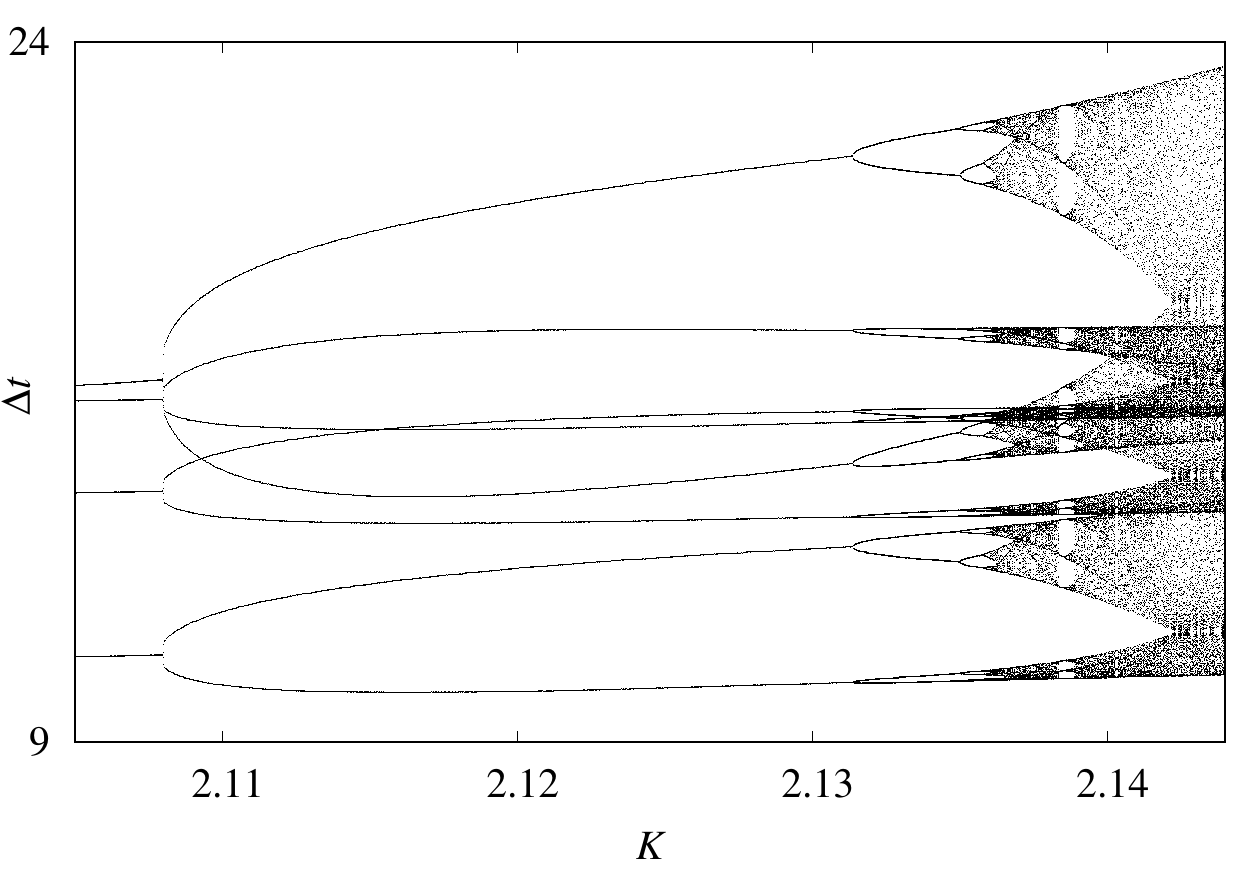}
\end{center}
\caption{Enlarged region from the bifurcation diagram in Fig.~\ref{Fig:1},
containing a period doubling cascade of the periodic pattern given in Fig.~\ref{Fig:1}(d).}
\label{Figure:PerDoubling}
\end{figure}
%%%%%%%%%%%%%%%%%%%%%%%%%%%%%%%%%%%%%%%
The bifurcation diagram in Fig.~\ref{Fig:1} shows that
in the parameter interval $K\in [2.1,2.15]$ we can observe a period doubling cascade
which transforms the periodic pattern given in Fig.~\ref{Fig:1}(d) into a chaotic regime.
In Fig.~\ref{Figure:PerDoubling} we show an enlargement of the bifurcation diagram for this region.
The observed period doubling bifurcations subsequently change the symmetry type of the  periodic solutions.
The solutions on the  primary branch, given in Fig.~\ref{Fig:1}(d),
return to the same state after exactly four velocity peaks and an index shift by one, i.e.
$$
\Delta t_{s+4}=\Delta t_{s} \quad\mathrm{~and~}\quad k_{s+4}=k_s+1\:\:\:\mathrm{mod}\:\:\:N.
$$
According to~(\ref{eq:RPO}) we have in this case a relative periodic solutions
with period $P= \Delta t_1 + \Delta t_2 + \Delta t_3+ \Delta t_4 $ and $\sigma(k)= k+1\:\:\:\mathrm{mod}\:\:\:N$.
At each period doubling the space-time symmetry is changed in a way
such that both the number of velocity peaks and index shifts
which are necessary to reach the same state increases by a factor of two.
In this way, the regular succession in space of the velocity peaks
of the solution in Fig.~\ref{Fig:1}(d)
 $$
\{k_s\}_{s\in\mathbb N} = \{3,4,5,6,4,5,6,7,5,\dots\}
$$
remains unchanged, while the sequences of inter-peak intervals~$\Delta t_s$
become more and more irregular.

\subsection{Torus breakup}

\begin{figure}[ht]
\begin{center}
\includegraphics[height=6.5cm]{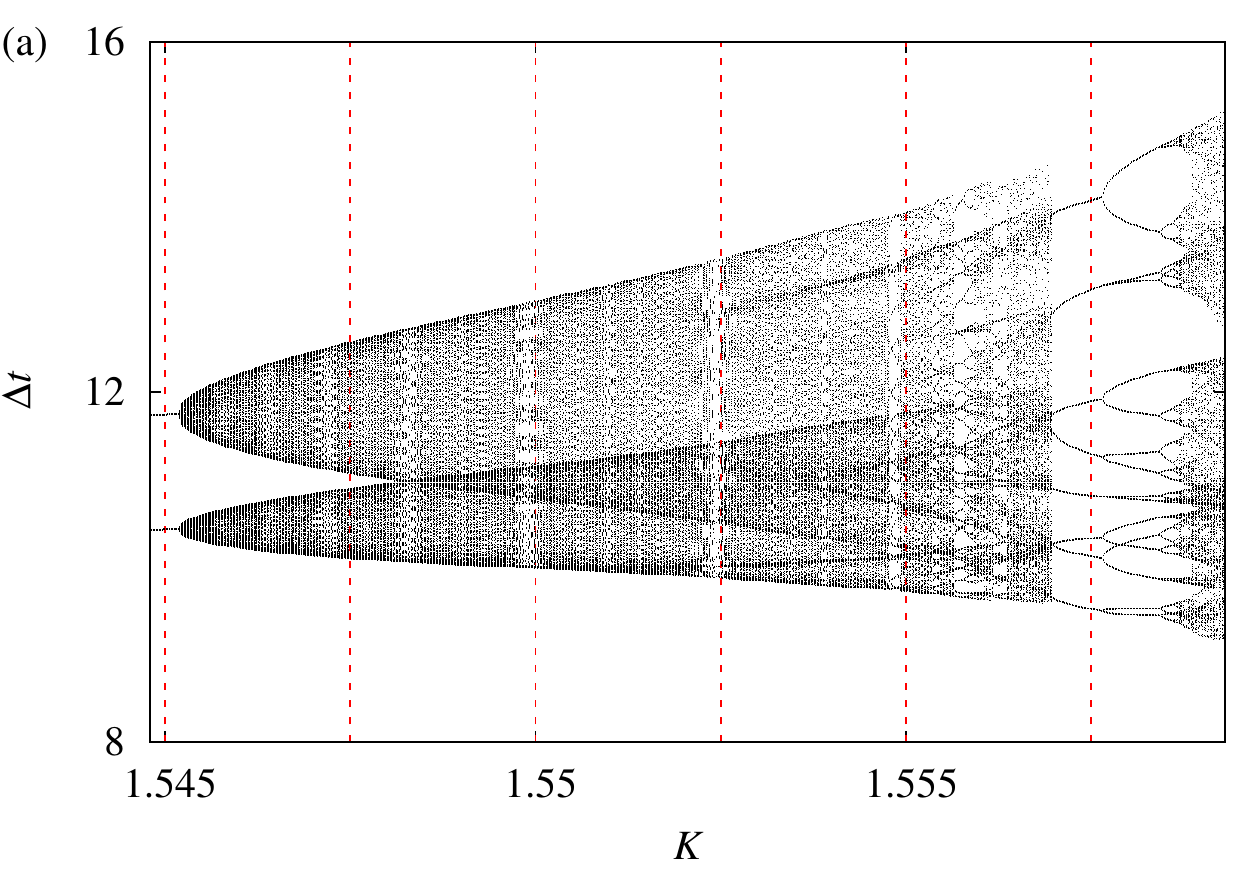}
\includegraphics[height=6.5cm]{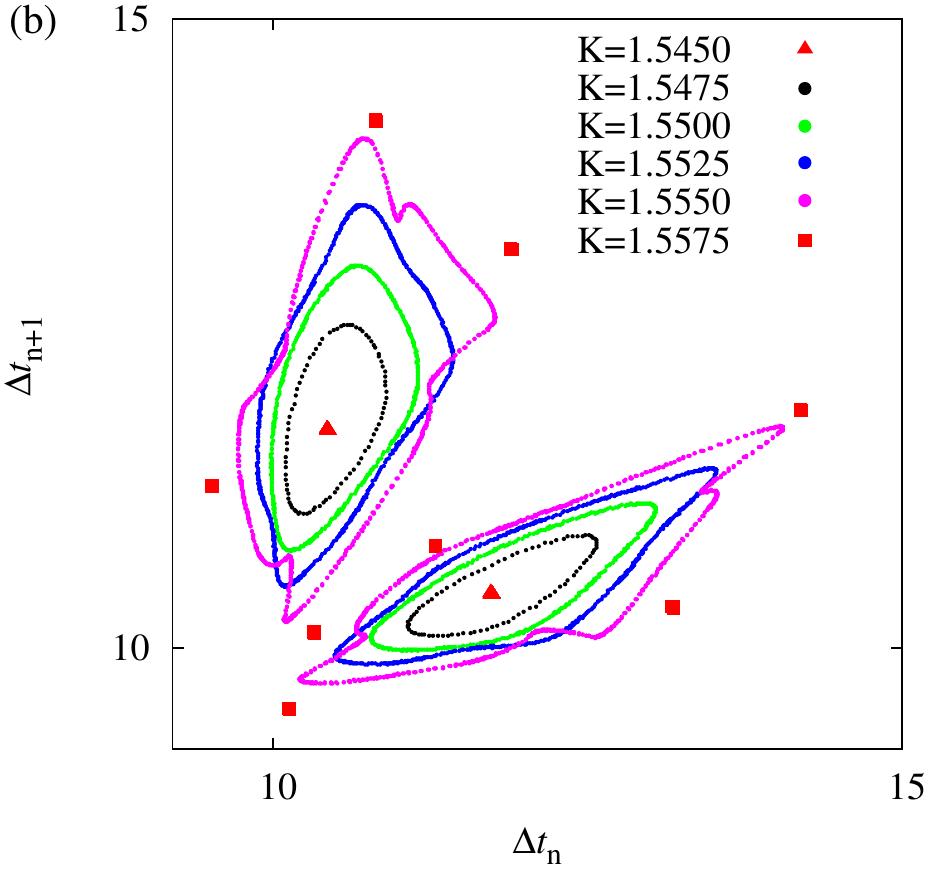}
\end{center}
\caption{(a) Enlarged region from the bifurcation diagram in Fig.~\ref{Fig:1},
containing a period torus bifurcation of the pattern Fig.~\ref{Fig:1}(a).
(b) Two-dimensional representation for selected values of~$K$ (dashed vertical lines in panel~(a)).}
\label{Figure:Torus}
\end{figure}

The periodic pattern given in Fig.~\ref{Fig:1}(a) loses its stability
in a torus bifurcation at $K\approx1.5452$ and a stable quasiperiodic pattern appears.
Fig.~\ref{Figure:Torus}(a) shows an enlargement of the corresponding region in the bifurcation diagram.
The two-dimensional representation in panel~(b) shows
the emergence of a stable invariant curve in the Poincar\'e section.
For further increasing parameter~$K$ there is a locking on the torus
and a subsequent transition to chaos via a torus breakup.
The primary pattern corresponds to a relative periodic orbit with period two in the Poincar\'e section.
Again, as in the case of the period doubling bifurcation,
the torus bifurcation does not change the pattern
given by the succession in space of the velocity peaks, which is in this case
$$
\{k_s\}_{s\in\mathbb N}=\{2,1,3,2,4,3,5,4,\dots\}
$$
while the sequences $\{\Delta t_{2s}\}_{s\in\mathbb N},\,\{\Delta t_{2s+1}\}_{s\in\mathbb N}$
of inter-peak intervals, which have been constant in the period two orbit,
start to vary periodically with an incommensurate period.
%%%%%%%%%%%%%%%%%%%%%%%%%%%%%%%%%%%%%%%
%%%%%%%%%%%%%%%%%%%%%%%%%%%%%%%%%%%%%%%

\subsection{Intermittency}

\begin{figure}[ht]
\begin{center}
\includegraphics[width=8.5cm]{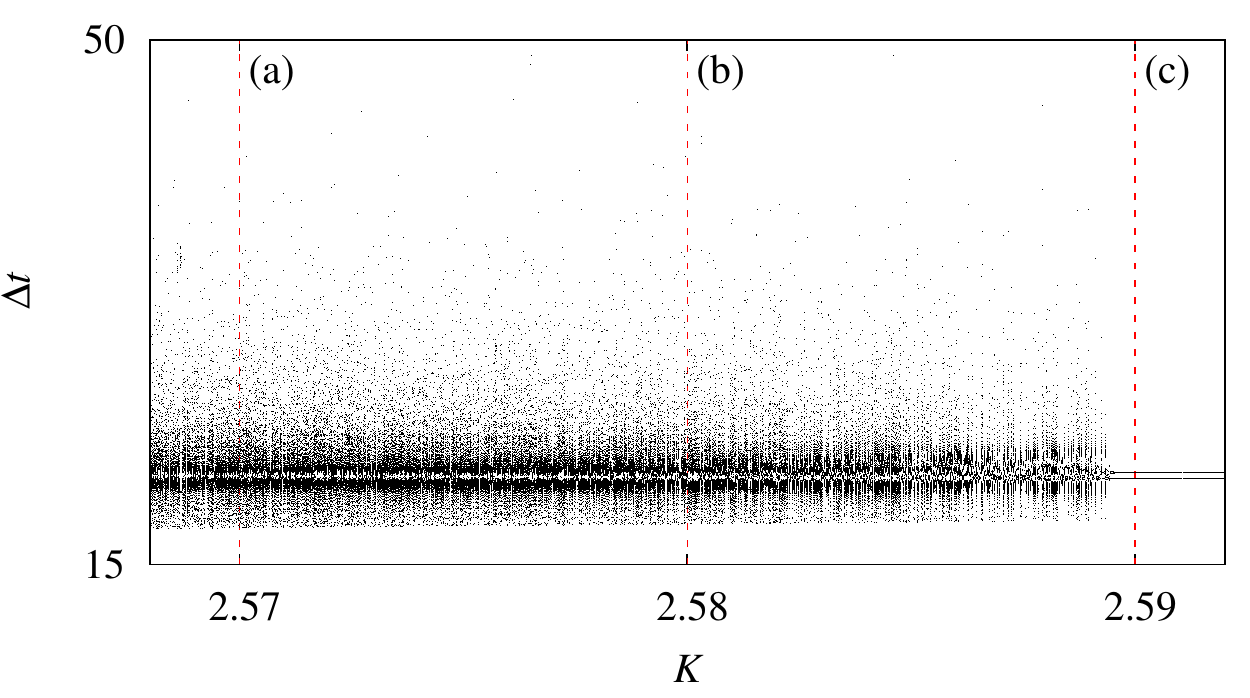}\\[2mm]
\includegraphics[width=8.5cm]{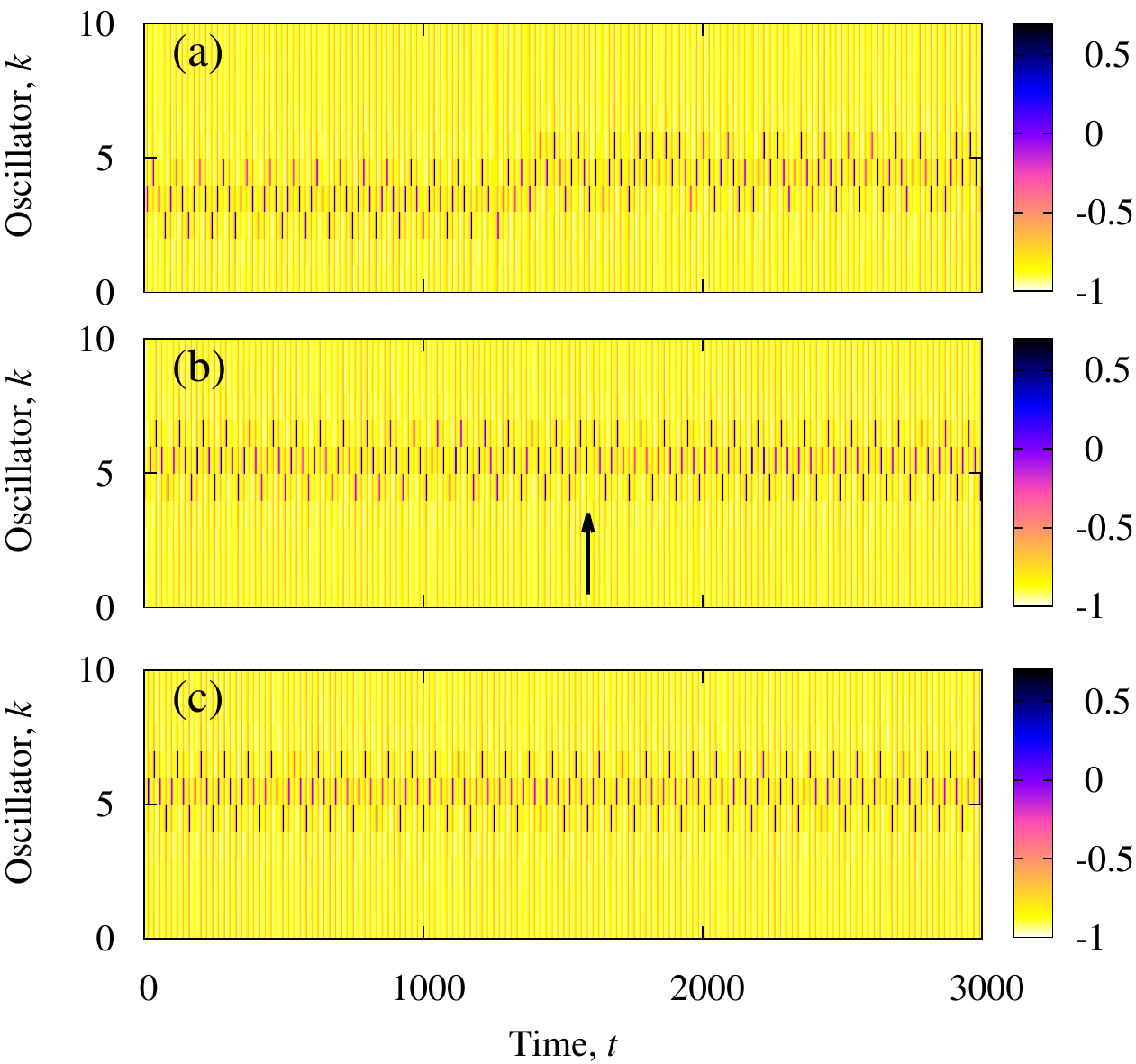}
\end{center}
\caption{Enlarged region from the bifurcation diagram in Fig.~\ref{Fig:1},
containing an intermittency transition from the pattern Fig.~\ref{Fig:1}(g) to chaos.}
\label{Figure:Intermittency}
\end{figure}

Within the complex bifurcation scenario in Fig.~\ref{Fig:1} one can also identify
a transition from regular to chaotic motion via intermittency.
Fig.~\ref{Figure:Intermittency} shows another enlargement,
displaying the region around $K_{\mathrm{crit}}\approx2.5894$
where for decreasing~$K$ the regular periodic pattern given in Fig.~\ref{Fig:1}(g)
loses its stability in an inverse (subcritical) period doubling bifurcation.
As we see in space-time plots in Fig.~\ref{Figure:Intermittency}(a)--(c),
this results in an intermittent behavior:
After the destabilization one can observe intervals of nearly periodic motion
that are interrupted by irregular occurring larger excursions in phase space.
According to general theory~\cite[Sec.~8.2]{Ott}, the average time interval between these excursions
scales like $(K-K_{\mathrm{crit}})^{-1}$.
Note that during the 3000 time units displayed in panel~(b)
there is only a single defect in the nearly regular pattern at $t\approx1590$ (see arrow).
This corresponds to an excursion from the period two orbit
that has been destabilized at the bifurcation
and is accompanied by large variations of the inter-peak-intervals.
In between the defects the trajectory stays close to the unstable period two orbit
and shows nearly regular inter-peak-intervals.
More distant to the bifurcation we observe more frequent excursions
from the regular motion, see panel~(a).
%%%%%%%%%%%%%%%%%%%%%%%%%%%%%%%%%%%%%%%
%%%%%%%%%%%%%%%%%%%%%%%%%%%%%%%%%%%%%%%

\subsection{Influence of the system size}
%%%%%%%%%%%%%%%%%%%%%%%%%%%%%%%%%%%%%%%
\begin{figure}[p]
\begin{center}
\includegraphics[width=0.99\textwidth]{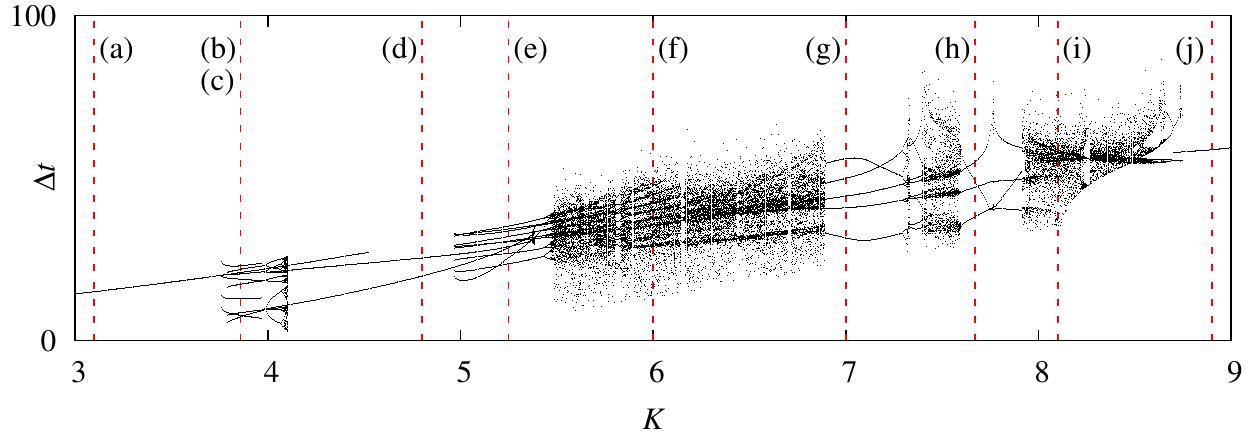}\\[2mm]
\includegraphics[width=0.99\textwidth]{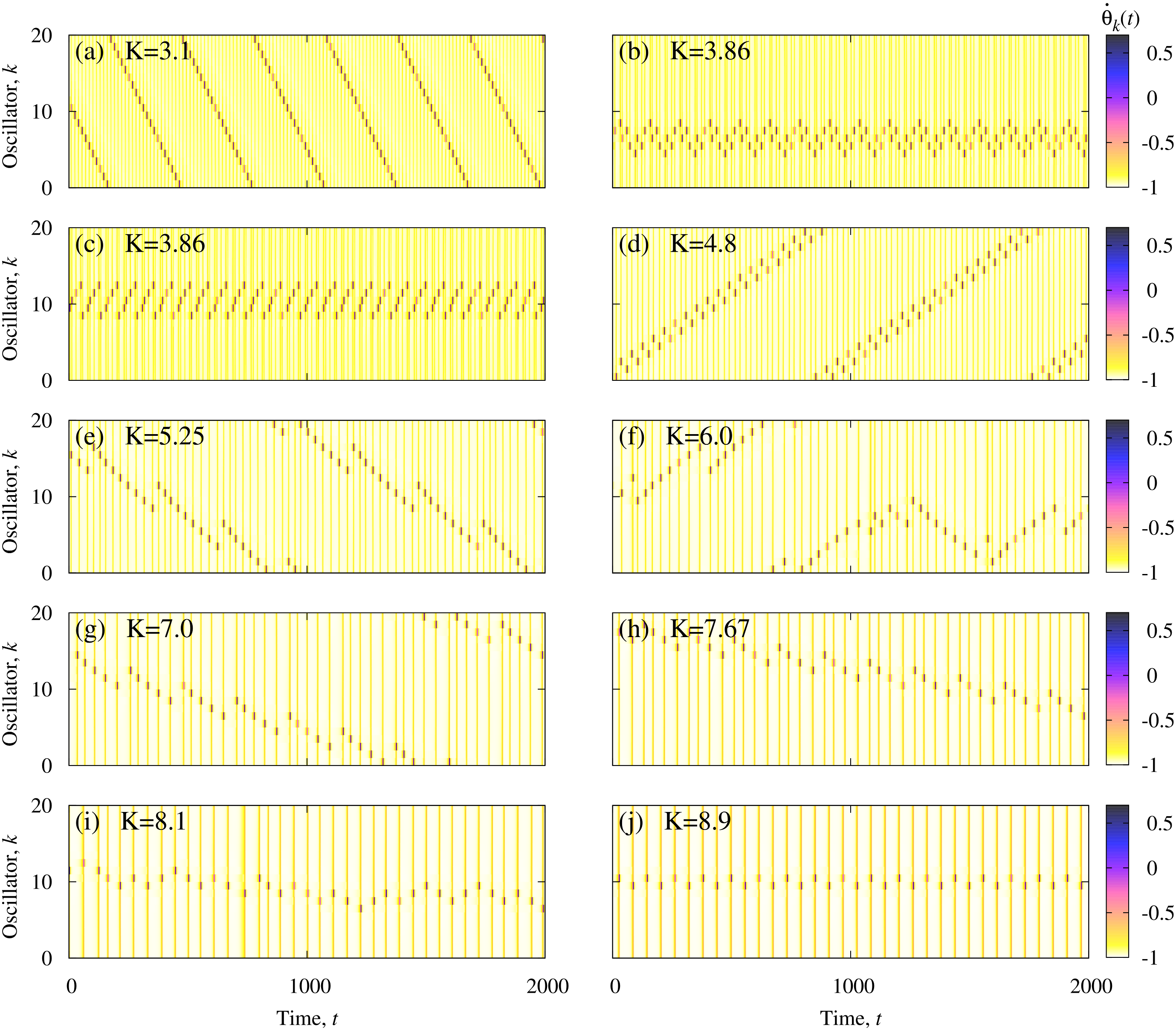}
\end{center}
\caption{Bifurcation diagram and space time plots
of the phase velocities for selected values of~$K$
(dashed vertical lines in the bifurcation diagram) for $N=20$ oscillators and $A = 0.9$.}
\label{Figure:BifDiagramN20}
\end{figure}
\begin{figure}[ht]
\begin{center}
\includegraphics[width=0.5\textwidth]{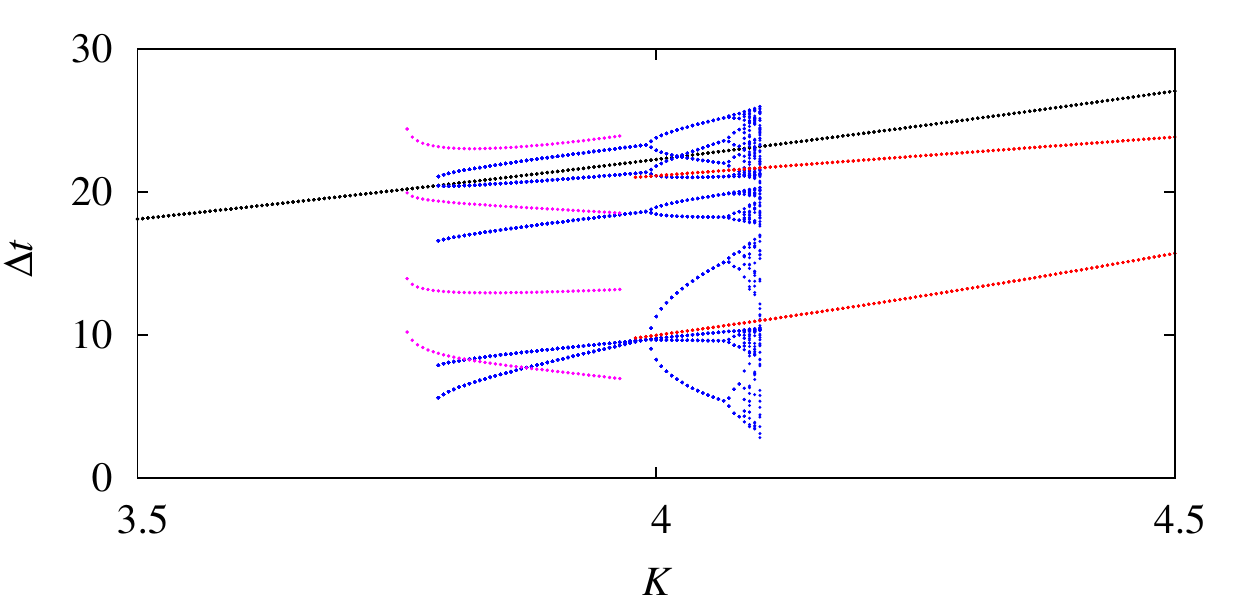}
\end{center}
\caption{Enlarged region from the bifurcation diagram in Fig.~\ref{Figure:BifDiagramN20},
containing four solution branches indicated by different colors, of which at least three overlap.}
\label{Figure:BifDiagramN20_enlarged}
\end{figure}

%%%%%%%%%%%%%%%%%%%%%%%%%%%%%%%%%%%%%%%
In Fig.~\ref{Figure:BifDiagramN20} we show the results of our simulations for $N=20$ oscillators.
We observe a similar scenario, including traveling
and stationary regular patterns as well as various types of irregular patterns.
Moreover, using the same procedure with a randomly chosen initial condition
for each parameter value as it was described above for $N=10$,
we detected here some regions of coexistence of several stable regular patterns.
In Fig.~\ref{Figure:BifDiagramN20} we show the results of our simulations for $N=20$ oscillators.
Remarkably, the pattern shown in panel~(a) extends to parameter values around $K=3.9$
where additionally the patterns shown in panels~(b) and~(c) exist,
see also Fig.~\ref{Figure:BifDiagramN20_enlarged} with four solution branches in different colors.
Moreover, pattern~(d) extends to the region around $K=5$ where also pattern~(e) can be found.
This is in contrast to the case $N=10$, where with the same procedure
we observed only small regions of multistability
(e.g., near the transition between the regimes in Fig.~\ref{Fig:1}(g) and~(h)
the stable regimes overlap for a range of~$K$ of~$0.01$).
However, for sufficiently large values of~$K$
we observe for $N=20$ again the simple zigzag pattern as for $N=10$ as the only attractor.
Since for increasing~$K$ in general the distance between the velocity peaks grows,
it seems likely that for increasing~$K$ a condensation to regular patterns
can be expected also for larger~$N$, however with a possibly larger extent of multistability.
%%%%%%%%%%%%%%%%%%%%%%%%%%%%%%%%%%%%%%%
%%%%%%%%%%%%%%%%%%%%%%%%%%%%%%%%%%%%%%%

\section{Self-modulated excitability}
Note that for increasing~$K$ one observes
a decreasing number of localized velocity peaks even for larger~$N$,
coming close to the completely coherent state.
For this type of behavior the continuum limit
describing chimera states by time independent averaged quantities
can be no longer considered as a valid description.
Instead, we suggest here a description of the patterns as
self-localized regions of excitation in a discrete excitable medium.
Indeed, the behavior of the velocity peaks
is often reminiscent to propagating excitation waves,
where preceding peaks trigger subsequent peaks in their neighborhood
and where the nonlocal nature of the coupling
can induce jumps and changes in their direction of propagation
in a regular or irregular manner.

To unveil the nature of system~(\ref{eq:1}) as a discrete excitable medium, we rewrite it as
\begin{equation}\label{eq:6}
 \frac{d \theta_k}{d t} = \omega - R_k \sin( \theta_k - \Theta_k + \alpha)
\mbox{,}\quad k=1,\ldots,N,
\end{equation}
using the phase~$\Theta_k$ and the absolute value~$R_k$
of the complex local mean field defined in~(\ref{eq:mf}).
For the completely coherent state with $\theta_1=\ldots=\theta_N$
we get the uniform rotation $\theta_j(t)=\Omega t$ with phase velocity $\Omega=\omega-\sin\alpha$.
Transforming equation~(\ref{eq:6}) into a corotating frame $\psi_k=\theta_k -\Omega t$
and inserting~(\ref{eq:3})  for the feedback, we obtain
\begin{align}
  \frac{d \psi_k}{d t} &= \sin\alpha - R_k \sin( \psi_k - \Psi_k + \alpha)\nonumber\\
  &=\cos(K(1-r))-R_k\cos(\psi_k-\Psi_k-K(1-r))\mbox{,}\quad k=1,\ldots,N.
  \label{eq:act:inh}
\end{align}
Here, $\Psi_k$ is the phase of the local mean field in the rotating coordinates~$\psi_k$.
The Jacobian of~(\ref{eq:act:inh}) with respect to $(\psi_1,\ldots,\psi_N)$ is zero
in the completely coherent state $\psi_1=\ldots\psi_N=0$,
where we have also $\Psi_1=\ldots=\Psi_N=0$, $R_1=\ldots R_N=1$, $r=1$.
Thus, the completely coherent state is a degenerate equilibrium,
which displays a saddle-node-on-limit-cycle bifurcation in each component.

During the intervals between the velocity peaks, the system comes
close to the uniformly phase locked solution and we have~$R_k(t)$
close below~$1$, and~$r(t)$ close below~$1$.
Regarding these quantities as external parameters, each oscillator~$k$
can be either in a stable excitable or oscillatory regime, depending on whether
$$
E_k = \frac{\cos(K(1-r))}{R_k}
$$
is smaller or bigger than~$1$, respectively.
The oscillator may change between these regimes
upon small changes of~$r$ and~$R_k$, which are both close below~$1$.
Note that an increasing total number of velocity peaks decreases
the global order parameter~$r$, thus decreasing the numerator of~$E_k$.
This acts as a global, i.e. long range inhibition,
taking all oscillators towards the non-oscillatory regime.
In particular, for increasing~$K$ we have a stronger inhibition,
which explains the increasing distance of the peaks in this situation.

In contrast, a locally increasing number of velocity peaks
(say, for oscillators~$j$ near a given index~$k$)
decreases the local order parameter~$R_k$ in the denominator,
increasing~$E_k$ and pushing the oscillator~$k$ towards the oscillatory regime.
Hence, it acts as a local activation. This interplay of activation and inhibition
in a discrete excitable medium close to threshold explains
a mechanism of self-modulated excitability,
where in the self-organized region of oscillatory motion
the observed variety of regular and irregular patterns can take place.

\section{Conclusions}
We have shown that in addition to the well elaborated approach
of studying chimera states in the framework of the continuum limit $N\to\infty$,
there is also a way to study their emergence for small~$N$
by methods of classical dynamical systems theory.
In particular, we could show that in the feedback system investigated here
chaotic chimera states can be found as true chaotic attractors,
emerging in well understood transitions such as period doubling cascades, torus breakup, and intermittency.
An increasing global feedback parameter~$K$,
pushing the dynamics closer towards the unstable completely coherent state
induces a condensation of the irregular chimera states
to simple regular patterns composed of localized phase slipping events of single oscillators.
Moreover, we could show that discrete excitable media
with nonlocal and global coupling can support not only the classical scenario of propagating of excitation waves,
but also a pattern formation process leading to self-localized regions of excitation
organized by an interplay between short range activation and long range inhibition.

\section{Acknowledgment}
We want to thank Y. L. Maistrenko for pointing out to us the importance of small size chimera states.
M.W. acknowledges the support by DFG within the collaborative research center SFB 910.


\begin{thebibliography}{21}
\expandafter\ifx\csname natexlab\endcsname\relax\def\natexlab#1{#1}\fi
\expandafter\ifx\csname bibnamefont\endcsname\relax
  \def\bibnamefont#1{#1}\fi
\expandafter\ifx\csname bibfnamefont\endcsname\relax
  \def\bibfnamefont#1{#1}\fi
\expandafter\ifx\csname citenamefont\endcsname\relax
  \def\citenamefont#1{#1}\fi
\expandafter\ifx\csname url\endcsname\relax
  \def\url#1{\texttt{#1}}\fi
\expandafter\ifx\csname urlprefix\endcsname\relax\def\urlprefix{URL }\fi
\providecommand{\bibinfo}[2]{#2}
\providecommand{\eprint}[2][]{\url{#2}}

\bibitem{kb2002}
\bibinfo{author}{\bibfnamefont{Y.}~\bibnamefont{Kuramoto}} \bibnamefont{and}
  \bibinfo{author}{\bibfnamefont{D.}~\bibnamefont{Battogtokh}},
  \bibinfo{journal}{Nonlinear Phenom. Complex Syst.}
  \textbf{\bibinfo{volume}{5}}, \bibinfo{pages}{380} (\bibinfo{year}{2002}).

\bibitem{wo2011}
\bibinfo{author}{\bibfnamefont{M.}~\bibnamefont{Wolfrum}} \bibnamefont{and}
  \bibinfo{author}{\bibfnamefont{O.~E.} \bibnamefont{Omel'chenko}},
  \bibinfo{journal}{Phys. Rev. E} \textbf{\bibinfo{volume}{84}},
  \bibinfo{pages}{015201} (\bibinfo{year}{2011}).

\bibitem{sow2014}
\bibinfo{author}{\bibfnamefont{J.}~\bibnamefont{Sieber}},
  \bibinfo{author}{\bibfnamefont{O.~E.} \bibnamefont{Omel'chenko}},
  \bibnamefont{and} \bibinfo{author}{\bibfnamefont{M.}~\bibnamefont{Wolfrum}},
  \bibinfo{journal}{Phys. Rev. Lett.} \textbf{\bibinfo{volume}{112}},
  \bibinfo{pages}{054102} (\bibinfo{year}{2014}).

\bibitem{as2004}
\bibinfo{author}{\bibfnamefont{D.~M.} \bibnamefont{Abrams}} \bibnamefont{and}
  \bibinfo{author}{\bibfnamefont{S.~H.} \bibnamefont{Strogatz}},
  \bibinfo{journal}{Phys. Rev. Lett.} \textbf{\bibinfo{volume}{93}},
  \bibinfo{pages}{174102} (\bibinfo{year}{2004}).

\bibitem{pa2015}
\bibinfo{author}{\bibfnamefont{M.~J.} \bibnamefont{Panaggio}} \bibnamefont{and}
  \bibinfo{author}{\bibfnamefont{D.~M.} \bibnamefont{Abrams}},
  \bibinfo{journal}{Nonlinearity} \textbf{\bibinfo{volume}{28}},
  \bibinfo{pages}{R67} (\bibinfo{year}{2015}).

\bibitem{tns2012}
\bibinfo{author}{\bibfnamefont{A.~F.} \bibnamefont{Taylor}},
  \bibinfo{author}{\bibfnamefont{S.}~\bibnamefont{Nkomo}}, \bibnamefont{and}
  \bibinfo{author}{\bibfnamefont{K.}~\bibnamefont{Showalter}},
  \bibinfo{journal}{Nature Physics} \textbf{\bibinfo{volume}{8}},
  \bibinfo{pages}{662} (\bibinfo{year}{2012}).

\bibitem{wk2013}
\bibinfo{author}{\bibfnamefont{M.}~\bibnamefont{Wickramasinghe}}
  \bibnamefont{and} \bibinfo{author}{\bibfnamefont{I.~Z.} \bibnamefont{Kiss}},
  \bibinfo{journal}{PLoS ONE} \textbf{\bibinfo{volume}{8}},
  \bibinfo{pages}{e80586} (\bibinfo{year}{2013}).

\bibitem{wk2014}
\bibinfo{author}{\bibfnamefont{M.}~\bibnamefont{Wickramasinghe}}
  \bibnamefont{and} \bibinfo{author}{\bibfnamefont{I.~Z.} \bibnamefont{Kiss}},
  \bibinfo{journal}{Phys. Chem. Chem. Phys.} \textbf{\bibinfo{volume}{16}},
  \bibinfo{pages}{18360} (\bibinfo{year}{2014}).

\bibitem{sskg-m2014}
\bibinfo{author}{\bibfnamefont{L.}~\bibnamefont{Schmidt}},
  \bibinfo{author}{\bibfnamefont{K.}~\bibnamefont{Sch{\"o}nleber}},
  \bibinfo{author}{\bibfnamefont{K.}~\bibnamefont{Krischer}}, \bibnamefont{and}
  \bibinfo{author}{\bibfnamefont{V.}~\bibnamefont{Garci{\'a}-Morales}},
  \bibinfo{journal}{Chaos} \textbf{\bibinfo{volume}{24}},
  \bibinfo{pages}{013102} (\bibinfo{year}{2014}).

\bibitem{hmrhos2012}
\bibinfo{author}{\bibfnamefont{A.~M.} \bibnamefont{Hagerstrom}},
  \bibinfo{author}{\bibfnamefont{T.~E.} \bibnamefont{Murphy}},
  \bibinfo{author}{\bibfnamefont{R.}~\bibnamefont{Roy}},
  \bibinfo{author}{\bibfnamefont{P.}~\bibnamefont{H{\"o}vel}},
  \bibinfo{author}{\bibfnamefont{I.}~\bibnamefont{Omelchenko}},
  \bibnamefont{and}
  \bibinfo{author}{\bibfnamefont{E.}~\bibnamefont{Sch{\"o}ll}},
  \bibinfo{journal}{Nature Physics} \textbf{\bibinfo{volume}{8}},
  \bibinfo{pages}{658} (\bibinfo{year}{2012}).

\bibitem{rrhsg2014}
\bibinfo{author}{\bibfnamefont{D.~P.} \bibnamefont{Rosin}},
  \bibinfo{author}{\bibfnamefont{D.}~\bibnamefont{Rontani}},
  \bibinfo{author}{\bibfnamefont{N.~D.} \bibnamefont{Haynes}},
  \bibinfo{author}{\bibfnamefont{E.}~\bibnamefont{Sch{\"o}ll}},
  \bibnamefont{and} \bibinfo{author}{\bibfnamefont{D.~J.}
  \bibnamefont{Gauthier}}, \bibinfo{journal}{Phys. Rev. E}
  \textbf{\bibinfo{volume}{90}}, \bibinfo{pages}{030902(R)}
  (\bibinfo{year}{2014}).

\bibitem{bzsl2015}
\bibinfo{author}{\bibfnamefont{F.}~\bibnamefont{B{\"o}hm}},
  \bibinfo{author}{\bibfnamefont{A.}~\bibnamefont{Zakharova}},
  \bibinfo{author}{\bibfnamefont{E.}~\bibnamefont{Sch{\"o}ll}},
  \bibnamefont{and}
  \bibinfo{author}{\bibfnamefont{K.}~\bibnamefont{L{\"u}dge}},
  \bibinfo{journal}{Phys. Rev. E}  (\bibinfo{year}{2015}).

\bibitem{mtfh2013}
\bibinfo{author}{\bibfnamefont{E.~A.} \bibnamefont{Martens}},
  \bibinfo{author}{\bibfnamefont{S.}~\bibnamefont{Thutupalli}},
  \bibinfo{author}{\bibfnamefont{A.}~\bibnamefont{Fourriere}},
  \bibnamefont{and}
  \bibinfo{author}{\bibfnamefont{O.}~\bibnamefont{Hallatschek}},
  \bibinfo{journal}{PNAS} \textbf{\bibinfo{volume}{110}},
  \bibinfo{pages}{10563} (\bibinfo{year}{2013}).

\bibitem{lc2001}
\bibinfo{author}{\bibfnamefont{C.~R.} \bibnamefont{Laing}} \bibnamefont{and}
  \bibinfo{author}{\bibfnamefont{C.~C.} \bibnamefont{Chow}},
  \bibinfo{journal}{Neural Computation} \textbf{\bibinfo{volume}{13}},
  \bibinfo{pages}{1473} (\bibinfo{year}{2001}).

\bibitem{oohs2013}
\bibinfo{author}{\bibfnamefont{I.}~\bibnamefont{Omelchenko}},
  \bibinfo{author}{\bibfnamefont{O.~E.} \bibnamefont{Omel'chenko}},
  \bibinfo{author}{\bibfnamefont{P.}~\bibnamefont{H{\"o}vel}},
  \bibnamefont{and}
  \bibinfo{author}{\bibfnamefont{E.}~\bibnamefont{Sch{\"o}ll}},
  \bibinfo{journal}{Phys. Rev. Lett.} \textbf{\bibinfo{volume}{110}},
  \bibinfo{pages}{224101} (\bibinfo{year}{2013}).

\bibitem{pm2014}
\bibinfo{author}{\bibfnamefont{D.}~\bibnamefont{Paz{\'o}}} \bibnamefont{and}
  \bibinfo{author}{\bibfnamefont{E.}~\bibnamefont{Montbri{\'o}}},
  \bibinfo{journal}{Phys. Rev. X} \textbf{\bibinfo{volume}{4}},
  \bibinfo{pages}{011009} (\bibinfo{year}{2014}).

\bibitem{l2014}
\bibinfo{author}{\bibfnamefont{C.~R.} \bibnamefont{Laing}},
  \bibinfo{journal}{Phys. Rev. E} \textbf{\bibinfo{volume}{90}},
  \bibinfo{pages}{010901(R)} (\bibinfo{year}{2014}).

\bibitem{gbcfmf2014}
\bibinfo{author}{\bibfnamefont{L.~V.} \bibnamefont{Gambuzza}},
  \bibinfo{author}{\bibfnamefont{A.}~\bibnamefont{Buscarino}},
  \bibinfo{author}{\bibfnamefont{S.}~\bibnamefont{Chessari}},
  \bibinfo{author}{\bibfnamefont{L.}~\bibnamefont{Fortuna}},
  \bibinfo{author}{\bibfnamefont{R.}~\bibnamefont{Meucci}}, \bibnamefont{and}
  \bibinfo{author}{\bibfnamefont{M.}~\bibnamefont{Frasca}},
  \bibinfo{journal}{Phys. Rev. E} \textbf{\bibinfo{volume}{90}},
  \bibinfo{pages}{032905} (\bibinfo{year}{2014}).

\bibitem{ab2015}
\bibinfo{author}{\bibfnamefont{P.}~\bibnamefont{Ashwin}} \bibnamefont{and}
  \bibinfo{author}{\bibfnamefont{O.}~\bibnamefont{Burylko}},
  \bibinfo{journal}{Chaos} \textbf{\bibinfo{volume}{25}},
  \bibinfo{pages}{013106} (\bibinfo{year}{2015}).

\bibitem{owm2010}
\bibinfo{author}{\bibfnamefont{O.~E.} \bibnamefont{Omel'chenko}},
  \bibinfo{author}{\bibfnamefont{M.}~\bibnamefont{Wolfrum}}, \bibnamefont{and}
  \bibinfo{author}{\bibfnamefont{Y.~L.} \bibnamefont{Maistrenko}},
  \bibinfo{journal}{Phys. Rev. E} \textbf{\bibinfo{volume}{81}},
  \bibinfo{pages}{065201} (\bibinfo{year}{2010}).

\bibitem{Ott}
\bibinfo{author}{\bibfnamefont{E.}~\bibnamefont{Ott}},
  \emph{\bibinfo{title}{Chaos in dynamical systems}}
  (\bibinfo{publisher}{Cambridge University Press},
  \bibinfo{address}{Cambridge}, \bibinfo{year}{2002}).

\end{thebibliography}
\end{document}